\pdfoutput=1
\documentclass[12pt]{article}

\newcommand{\paragraphNoSkip}[1]{\noindent\textbf{#1.}}
\providecommand{\keywords}[1]{\noindent\textbf{{Keywords:}} #1.}
\usepackage[letterpaper, margin=1.2in]{geometry}
\usepackage[backend=biber, style=numeric, sorting=nyt, maxbibnames=10, maxalphanames=10, minalphanames=10]{biblatex}

\let\localcite\cite
\renewcommand{\cite}[1]{\localcite{#1}}
\let\localciteauthor\citeauthor
\newcommand{\citet}[1]{\localciteauthor{#1}~\localcite{#1}}

\usepackage[T1]{fontenc}
\usepackage[english]{babel}
\usepackage{csquotes}

\usepackage{mathtools}

\usepackage{amsthm}

\newtheorem*{example*}{Example}

\theoremstyle{remark}

\usepackage{url}
\PassOptionsToPackage{hyphens}{url}

\usepackage{hyperref}
\PassOptionsToPackage{breaklinks}{hyperref}

\usepackage{xcolor}
\hypersetup{colorlinks, linkcolor={red!50!black}, citecolor={blue!50!black}, urlcolor={blue!80!black}}

\usepackage{enumitem}

\usepackage[capitalise]{cleveref}
\crefname{definition}{Definition}{Definitions}
\crefname{example}{Example}{Examples}
\crefname{remark}{Remark}{Remarks}

\usepackage{tikz}
\usetikzlibrary{external} 
\tikzset{external/system call={%
        pdflatex \tikzexternalcheckshellescape
        -shell-escape -interaction=batchmode -output-directory=figures/ 
        -jobname "\image" "\texsource"}}
\usepackage{pgfplots}
\usetikzlibrary{pgfplots.dateplot, shapes.misc}
\usepgfplotslibrary{colorbrewer}
\pgfplotsset{compat=newest}
\makeatletter\newcommand*\short[1]{\expandafter\@gobbletwo\number\numexpr#1\relax}\makeatother

\usepackage{booktabs}
\usepackage{xspace}
\usepackage{caption}

\usepackage[symbols, acronym, nonumberlist, nogroupskip, stylemods={mcols,longbooktabs}, section=subsection, numberedsection]{glossaries-extra}
\makenoidxglossaries
\glssetcategoryattribute{acronym}{nohyper}{true}
\setabbreviationstyle[acronym]{long-short}

\newacronym{PoliFi}{PoliFi}{political finance}
\newacronym{DeFi}{DeFi}{decentralized finance}
\newacronym{DEX}{DEX}{decentralized exchange}
\newacronym{TVL}{TVL}{total value locked}
\newacronym{MEV}{MEV}{miner-extractable value}
\newacronym{PoS}{PoS}{proof-of-stake}
\newacronym{GENIUS}{GENIUS}{Guiding and Establishing National Innovation for U.S. Stablecoins}
\newacronym{ANOVA}{ANOVA}{analysis of variance}
\newacronym{PELT}{PELT}{Pruned Exact Linear Time}
\newacronym{CAR}{CAR}{cumulative abnormal returns}
\newacronym{HAC}{HAC}{heteroskedasticity- and autocorrelation-consistent}
\newacronym{OLS}{OLS}{ordinary least squares}
\newacronym{ADF}{ADF}{Augmented Dickey Fuller}
\newacronym{iff}{iff}{if and only if}
\newacronym{AIC}{AIC}{Akaike information criterion}
\newacronym{ToS}{ToS}{terms of service}
\newacronym{ICO}{ICO}{initial coin offering}
\newacronym{KPSS}{KPSS}{Kwiatkowski, Phillips, Schmidt, and Shin}

\glsxtrnewsymbol[description={The \$TRUMP token.}]{trump}{\ensuremath{\text{\$TRUMP}}}
\newcommand{\trump}{{\gls[hyper=false]{trump}}\xspace}

\glsxtrnewsymbol[description={The \$MELANIA token.}]{melania}{\ensuremath{\text{\$MELANIA}}}
\newcommand{\melania}{{\gls[hyper=false]{melania}}\xspace}

\begin{document}
\title{PoliFi Tokens and the Trump Effect}

\author{Ignacy Nieweglowski$^{1,*}$ \and Aviv Yaish$^{2,*}$ \and Fahad Saleh$^{3}$ \and Fan Zhang$^2$}
\date{\small{$^1$Staples High School, $^2$Yale University and IC3, $^3$University of Florida}}
\maketitle

\begin{abstract}
Cryptoassets launched by political figures, e.g., \gls{PoliFi} tokens, have recently attracted attention.
Chief among them are the eponymous tokens backed by the 47th president and first lady of the United States, \trump and \melania.
We empirically analyze both, and study their impact on the broad \gls{DeFi} ecosystem.
Via a comparative longitudinal study, we uncover a ``Trump Effect'':
the behavior of these tokens correlates positively with presidential approval ratings, whereas the same tight coupling does not extend to other cryptoassets and administrations.
We additionally quantify the ecosystemic impact, finding that the fervor surrounding the two assets was accompanied by capital flows towards associated platforms like the Solana blockchain, which also enjoyed record volumes and fee expenditure.

\end{abstract}

\section{Introduction}
Cryptoassets promise a decentralized economic order based on a cryptographic notion of trust \cite{nakamoto2008bitcoin}.
\Gls{DeFi} platforms extend this promise beyond mere peer-to-peer payments by offering a variety of economic services \cite{yaish2023suboptimality}.
These idealistic beginnings have given rise to many novel digital assets, including so-called \emph{memecoins}:
tokens commonly named after online memes or intended to serve some humorous purpose \cite{li2022will, zhang2021popular}.
These have recently taken a turn towards the political, exemplified by tokens backed by government officials and related figures, e.g., Donald J. Trump's \trump token.
We call such assets \emph{\glsxtrfull{PoliFi} tokens}.

As memecoins are often created with little to no technical innovation, we hypothesize that their economic behavior is sensitive to celebrity endorsements, current events, and broader market sentiment.
When considering political memecoins and the surrounding \gls{PoliFi} ecosystem, their economic performance may be affected by approval ratings and even government decisions pertaining to a-political cryptoassets.
Confirming this hypothesis would imply that political capital is not only measured by polls but also by the activity and prices of \gls{PoliFi} assets.
This leads to a natural question:

\begin{quote}
    \emph{Can we quantify the relationship between \gls{PoliFi} token behavior and the standing of the corresponding administrations?}
\end{quote}

\subsection{This Work}
We answer this question by focusing on two memecoins that have received broad public attention: \trump and \melania.

\tikzsetnextfilename{Timeline}
\begin{figure}
\centering
\begin{tikzpicture}[
    event/.style={circle, fill=gray!50, inner sep=2pt},
    highlight/.style={circle, fill=red!70, inner sep=2pt},
    label/.style={font=\footnotesize, align=left},
    scale=0.9, transform shape,
]
    \def\timelinestart{0}
    \def\timelineend{16}
    \def\timelineheight{0}

    \draw[thick] (\timelinestart,\timelineheight) -- (\timelineend,\timelineheight);

    \def\scale{0.056}
    
    \pgfmathsetmacro{\janend}{12*\scale}
    \pgfmathsetmacro{\febend}{\janend+28*\scale}
    \pgfmathsetmacro{\marend}{\febend+31*\scale}
    \pgfmathsetmacro{\aprend}{\marend+30*\scale}
    \pgfmathsetmacro{\mayend}{\aprend+31*\scale}
    \pgfmathsetmacro{\junend}{\mayend+30*\scale}
    \pgfmathsetmacro{\julend}{\junend+31*\scale}
    \pgfmathsetmacro{\augend}{\julend+31*\scale}
    \pgfmathsetmacro{\sepend}{\augend+30*\scale}
    \pgfmathsetmacro{\octend}{\sepend+31*\scale}

    \foreach \x/\month in {0/Jan 20, \janend/Feb, \febend/Mar, \marend/Apr, \aprend/May, \mayend/Jun, \junend/Jul, \julend/Aug, \augend/Sep, \sepend/Oct} {
        \draw[thick] (\x,0.15) -- (\x,-0.15);
        \node[below, font=\tiny] at (\x,-0.3) {\month};
    }

    \draw[thick] (\octend,0.15) -- (\octend,-0.15);
    \node[below, font=\tiny] at (\octend,-0.3) {Oct 31};

    \node[event] (e1) at (.672, 0) {};
    \node[label, above right, rotate=45, anchor=south west] at (.672, 0.1) {1st Tariffs};

    \node[event] (e4) at (4.08, 0) {};
    \node[label, above right, rotate=45, anchor=south west] at (3.9, 0.1) {Liberation Day};
    
    \node[event] (e5) at (5.208, 0) {};
    \node[label, above right, rotate=45, anchor=south west] at (5.3, 0.1) {Crypto Dinner\\Announced};

    \node[event] (e6) at (7.28, 0) {};
    \node[label, above right, rotate=45, anchor=south west] at (7.1, 0.1) {Musk Oval\\Office Farewell};

    \node[event] (e7) at (8.288, 0) {};
    \node[label, above right, rotate=45, anchor=south west] at (8.5, 0.1) {GENIUS Act\\Passed};

    \node[event] (e10) at (13.72, 0) {};
    \node[label, above right, rotate=45, anchor=south west] at (13.5, 0.1) {Kirk Memorial};

    \node[event] (e11) at (14.728, 0) {};
    \node[label, above right, rotate=45, anchor=south west] at (14.9, 0.1) {China Tariff\\Threats};
\end{tikzpicture}
  \vspace{-10pt}
  \caption{Some of the notable events covered in our study.}
  \vspace{-20pt}
  \label{fig:Timeline}
\end{figure}
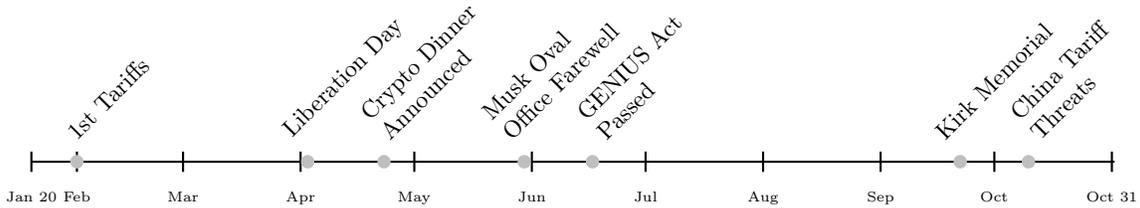

First, we analyze how both are impacted by political events.
We employ a ``forward-backward'' approach, using the \gls{PELT} change-point detection algorithm \cite{truong2020selective} to identify points where the statistical properties of time series data change, and then connect them to coinciding notable events.
Several events were identified, including those relating to US tariffs, the announcement of the ``crypto dinner'' to top \trump holders, and the advancement of crypto-friendly \gls{GENIUS} act, among others (see \cref{fig:Timeline}).

To study the relationship of \gls{PoliFi} assets with the corresponding figures' public reception, we thoroughly analyze the correlation of approval ratings and token behavior.
The data suggests a \emph{Trump effect}: the behavior of \trump and \melania correlates with ratings to a larger extent than other tokens, while the general market tends to behave differently under Trump's second administration compared to preceding ones.

We proceed to analyze the interplay of \gls{PoliFi} and the broad ecosystem, considering both the impact on Solana, the blockchain on which \trump and \melania launched, and on others.
Thus, in the aftermath of \trump's launch, Solana's market cap and \gls{DEX} volume increased at the expense of competitors'.

To perform our analysis, we assemble a corpus from several sources, including cryptoasset market data, presidential approval ratings, social media posts by political figures, and headlines from notable news sources.
In total, we create a comprehensive timeline of the second Trump administration's actions and their aftermath.
As a service to the community and to enable future research, upon publication, we will release comprehensive instructions for constructing our corpus from public sources while adhering to the associated \gls{ToS}.

\section{Related Work}
\label{sec:RelatedWork}

\paragraphNoSkip{Memecoins}
Memecoins have been studied empirically before.
\citet{zhang2021popular} find that positive shocks have a larger impact than negative ones on the volatility of a prominent memecoin, Dogecoin.
Focusing on broader market impact, \citet{li2022will} find that spikes in memecoin value tend to result in a crash of the cryptoasset market.

\paragraphNoSkip{Token Endorsement}
The impact of celebrity endorsement on cryptoassets is analyzed by \citet{white2024effect}, focusing on blockchain-based platforms who raise funds by performing an \gls{ICO} of utility tokens (defined as tokens serving platform-designated purposes, like fee payment).
The authors find that such endorsements assist in meeting higher fundraising goals.

\paragraphNoSkip{\Glsxtrfull{PoliFi}}
\citet{long2025bridging} perform a multi-modal study by applying sentiment analysis to data sourced from online communities, and conclude that comments on politically-tinged memecoins have more stable sentiment, possibly reflecting targeted bot activity.
An empirical analysis of blockchain-based betting markets is carried out by \citet{chen2025political}, who attempt to predict political leanings from user bets, and evaluate markets' success in predicting the outcomes of political bets.

\paragraphNoSkip{Trump and Financial Markets}
\citet{benton2020does} focus on Trump's first administration, and perform an empirical analysis of how Trump's posts on Twitter (as it was then called) affected the USD-Mexican Peso exchange rate.
The authors find that the negative views expressed in these posts led to an increase in volatility.

\section{Methodology}

\subsubsection*{Data}
We compile a comprehensive dataset from multiple sources.

\paragraphNoSkip{Market Data}
Daily price and market capitalization for \trump, \melania, Bitcoin (BTC), Ethereum (ETH), Solana (SOL), Dogecoin (DOGE), and Binance's BNB Smart Chain (BNB) are sourced from CoinGecko.
Daily \gls{TVL}, transaction fees, and \gls{DEX} activity are obtained from DefiLlama.

\paragraphNoSkip{Political Data}
We assemble a corpus of off-chain political signals.
Using TruthBrush \cite{mccain2022truthbrush}, we scrape all posts by Donald and Melania Trump during the period under study.
For our longitudinal cross-administration study, we obtain Gallup's approval ratings for all US presidents whose tenure overlapped or came after the launch of the first major cryptoasset, BTC.
We construct a timeline of notable events that took place during Trump's second presidency (until the time of writing), sourced from ABC, AP News, CNN, CNBC, The New York Times, The Guardian, Forbes, and USA Today.

\subsubsection*{Formal Methods}
We employ several primary data analysis methods.

\paragraphNoSkip{Spearman Correlation}
We use the canonical Spearman's rank correlation coefficient, also known as Spearman's $\rho$, to evaluate monotone (not necessarily linear) dependence.
In the interest of space, we refer unfamiliar readers to \citet{cohen2014applied}.

\paragraphNoSkip{\Glsxtrfull{ADF} Test}
The \gls{ADF} test is a common approach to evaluating the stationarity and mean reversion of time series data \cite{dickey1979distribution, said1984testing, dolado2002fractional}.
A stationary time series exhibits constant statistical properties over time, more specifically, constant mean, variance, and autocovariance structure.
The \gls{ADF} test removes autocorrelation in input data, and accounts for trends in the series.
In particular, we follow the standard approach of specifying a linear trend and use the standard \gls{AIC} method for selecting the number of lagged values of the time series to consider \cite{cavanaugh2019akaike}.

\paragraphNoSkip{\gls{PELT}}
To align statistical changes with real-world events, we employ the \gls{PELT} algorithm \cite{truong2020selective}.
\gls{PELT} segments an $n$ point time series $y_{1:n}$ by finding a set of $K$ points $\tau = \left(\tau_1, \dots, \tau_K\right)$ in $y$ (i.e., given boundaries $\tau_0 = 0$ and $\tau_{K+1} = n$ then $\tau_i < \tau_j$ \gls{iff} $i < j$) minimizing a cost function $C$ plus a penalty $\beta$ (specifically, we use a strict penalty value of $\beta = 50$ coupled with the commonly-used $rbf$ cost function \cite{truong2020selective}):

$$
\min_{\{\tau_k\}} \left[
    \sum_{k=0}^{K} \mathcal{C} \big(
        x_{(\tau_k+1):\tau_{k+1}}
    \big) \;+\; \beta K
\right]
.
$$

\section{Results}
To answer our main question, we apply the above methodology to our dataset.
In particular, we identify two phenomena which, when combined, we call the \emph{Trump effect}; we soon describe both.
While our data covers \trump's existence from launch to the time of writing (roughly $10$ months), this still prevents making strong claims, e.g., with respect to persistent correlations or long-run efficiency.
Thus, the results are preliminary.
To ensure consistency between preceding administrations and the current one, we consider only the first $10$ months of data for each administration.

\subsection{Correlation Study}
\begin{figure*}
    \centering%
    \tikzsetnextfilename{CorrelationMatrix}%
\begin{minipage}{.49\textwidth}
\centering
\scalebox{0.49}{\begin{tikzpicture}
\begin{axis}[
  scale=0.8,
  width=14cm,
  height=5cm,
  enlargelimits=false,
  colormap={slategraywhite}{rgb255=(191,0,0), rgb255=(64,0,0),},
  colorbar,
  colorbar style={ylabel={Spearman},},
  xlabel=Token,
  ylabel=Administration,
  nodes near coords,
  nodes near coords align={center},
  every node near coord/.append style={text=white, font=\normalsize},
  xtick={BTC, DOGE, ETH, SOL, BNB, TRUMP, MELANIA},
  symbolic x coords={BTC, DOGE, ETH, SOL, BNB, TRUMP, MELANIA},
  ytick={Trump 2, Biden, Trump 1, Obama 2},
  symbolic y coords={Trump 2, Biden, Trump 1, Obama 2},
]
  \addplot [matrix plot, mesh/cols=7, point meta=explicit, /pgf/number format/fixed, /pgf/number format/precision=2,] table [col sep=comma, x=name, y=admin, z=r_returns, meta=r_returns] {figures/correlation_matrix.csv};
\end{axis}
\end{tikzpicture}}
\end{minipage}
\hfill
\tikzsetnextfilename{CorrelationApproval}%
\begin{minipage}{.49\textwidth}
\centering
\scalebox{0.49}{\begin{tikzpicture}
\begin{axis}[
  scale=0.8,
  width=14cm,
  height=5cm,
  enlargelimits=false,
  colormap={slategraywhite}{rgb255=(191,0,0), rgb255=(64,0,0),},
  colorbar,
  colorbar style={ylabel={Spearman},},
  xlabel=Token,
  ylabel=Administration,
  nodes near coords,
  nodes near coords align={center},
  every node near coord/.append style={text=white, font=\normalsize},
  xtick={BTC, DOGE, ETH, SOL, BNB, TRUMP, MELANIA},
  symbolic x coords={BTC, DOGE, ETH, SOL, BNB, TRUMP, MELANIA},
  ytick={Trump 2, Biden, Trump 1, Obama 2},
  symbolic y coords={Trump 2, Biden, Trump 1, Obama 2},
]
  \addplot [matrix plot, mesh/cols=7, point meta=explicit, /pgf/number format/fixed, /pgf/number format/precision=2,] table [col sep=comma, x=name, y=admin, z=r_price, meta=r_price] {figures/correlation_matrix_price.csv};
\end{axis}
\end{tikzpicture}}
\end{minipage}
\caption{The Spearman correlation between approval ratings and log-returns (left), and between ratings and prices (right).}
\label{fig:CorrelationApproval}
\end{figure*}
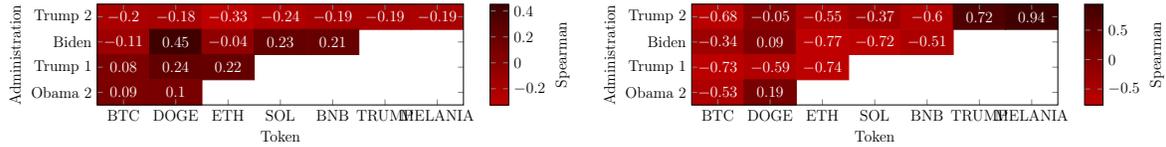

To tease out the Trump effect from the data, we calculate the Spearman rank correlation between the performance of multiple assets and presidential approval ratings as a proxy for political sentiment.
We look at price and returns respectively (see \cref{fig:CorrelationApproval}).
Throughout the analysis, each weekly approval data point is matched to the token data point corresponding to the same day.

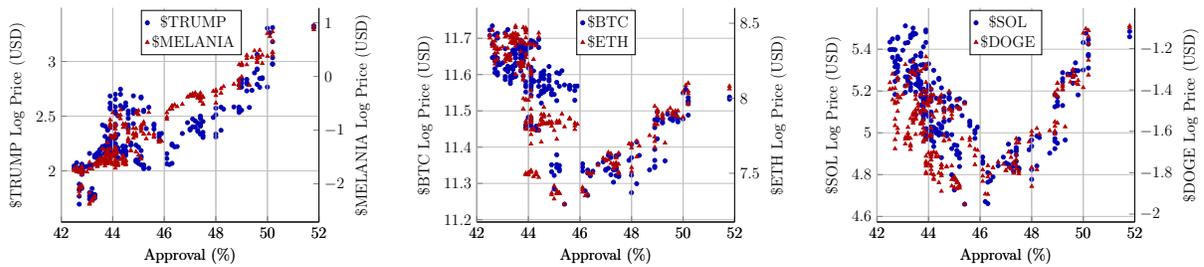
\begin{figure*}
\centering
  \tikzsetnextfilename{ApprovalScatterTrumpMelania}
  \begin{minipage}{.3\textwidth}
  \centering
  \scalebox{0.5}{\begin{tikzpicture}
    \begin{axis}[
    xlabel={Approval (\%)},
    ylabel={\trump Log Price (USD)},
    xmajorgrids,
    ymajorgrids,
    xmin=42,
    xmax=52,
    axis y line*=left,
    axis x line*=bottom,
    ]
    \addplot[only marks, mark=*, mark size=1.5pt, blue!70!black] table [col sep=comma, x=pct, y=log_TRUMP_price] {figures/TrumpT2_all_tokens.csv};
    \label{fig:ApprovalScatterTRUMP}
    \end{axis}
    \begin{axis}[
    xlabel={Approval (\%)},
    ylabel={\melania Log Price (USD)},
    xmajorgrids,
    xmin=42,
    xmax=52,
    axis y line*=right,
    axis x line*=bottom,
    legend style={at={(0.7,0.99)}},
    ]
    \addlegendimage{/pgfplots/refstyle=fig:ApprovalScatterTRUMP};
    \addlegendentry{\trump};
    
    \addplot[only marks, mark=triangle*, mark size=1.5pt, red!70!black] table [col sep=comma, x=pct, y=log_MELANIA_price] {figures/TrumpT2_all_tokens.csv};
    \addlegendentry{\melania};
    \end{axis}
\end{tikzpicture}}
\end{minipage}
\hfill
\tikzsetnextfilename{ApprovalScatterBTCETH}
\begin{minipage}{.3\textwidth}
\centering
\scalebox{0.5}{\begin{tikzpicture}
    \begin{axis}[
    xlabel={Approval (\%)},
    ylabel={\$BTC Log Price (USD)},
    xmajorgrids,
    ymajorgrids,
    xmin=42,
    xmax=52,
    axis y line*=left,
    axis x line*=bottom,
    ]
    \addplot[only marks, mark=*, mark size=1.5pt, blue!70!black] table [col sep=comma, x=pct, y=log_BTC_price] {figures/TrumpT2_all_tokens.csv};
    \label{fig:ApprovalScatterBTC}
    \end{axis}
    \begin{axis}[
    xlabel={Approval (\%)},
    ylabel={\$ETH Log Price (USD)},
    xmajorgrids,
    xmin=42,
    xmax=52,
    axis y line*=right,
    axis x line*=bottom,
    legend style={at={(0.615,0.99)}},
    ]
    \addlegendimage{/pgfplots/refstyle=fig:ApprovalScatterBTC};
    \addlegendentry{\$BTC};
      
    \addplot[only marks, mark=triangle*, mark size=1.5pt, red!70!black] table [col sep=comma, x=pct, y=log_ETH_price] {figures/TrumpT2_all_tokens.csv};
    \addlegendentry{\$ETH};
    \end{axis}
\end{tikzpicture}}
\end{minipage}
\hfill
\tikzsetnextfilename{ApprovalScatterSOLDOGE}
\begin{minipage}{.3\textwidth}
  \centering
  \scalebox{0.5}{\begin{tikzpicture}
    \begin{axis}[
    xlabel={Approval (\%)},
    ylabel={\$SOL Log Price (USD)},
    xmajorgrids,
    ymajorgrids,
    xmin=42,
    xmax=52,
    axis y line*=left,
    axis x line*=bottom,
    ]
    \addplot[only marks, mark=*, mark size=1.5pt, blue!70!black] table [col sep=comma, x=pct, y=log_SOL_price] {figures/TrumpT2_all_tokens.csv};
    \label{fig:ApprovalScatterSOL}
    \end{axis}
    \begin{axis}[
    xlabel={Approval (\%)},
    ylabel={\$DOGE Log Price (USD)},
    xmajorgrids,
    xmin=42,
    xmax=52,
    axis y line*=right,
    axis x line*=bottom,
    legend style={at={(0.64,0.99)}},
    ]
    \addlegendimage{/pgfplots/refstyle=fig:ApprovalScatterSOL};
    \addlegendentry{\$SOL};
    
    \addplot[only marks, mark=triangle*, mark size=1.5pt, red!70!black] table [col sep=comma, x=pct, y=log_DOGE_price] {figures/TrumpT2_all_tokens.csv};
    \addlegendentry{\$DOGE};
    \end{axis}
\end{tikzpicture}}
\end{minipage}
\caption{Across the board, log-prices correlate well with approval ratings greater than $46\%$. The \trump and \melania political memecoins stand out in continuing this correlation for lower ratings (left figure), while other tokens (e.g., \$BTC, center figure) and memecoins (e.g., \$DOGE, right figure) exhibit an opposite correlation.}
\label{fig:ApprovalScatter}
\end{figure*}

\paragraphNoSkip{Trump Effect I: \trump and \melania} \cref{fig:CorrelationApproval} (right) shows that \trump price exhibits a positive association with approval ($\rho=0.72$), as does \melania ($\rho=0.94$).
However, major non-\gls{PoliFi} tokens are negatively correlated, e.g., \$BTC ($\rho=-0.68$) and \$ETH ($\rho=-0.55$).
We investigate this further in \cref{fig:ApprovalScatter}, a scatterplot comparing token log-price datapoints with the corresponding approval ratings.
It becomes apparent that \trump and \melania log-prices have a relatively consistent positive correlation with ratings.
Meanwhile, other tokens have a negative correlation for ratings lower than $46\%$, and a positive one for higher ratings.

Besides memecoins (i.e., \$DOGE, \trump, and \melania), the negative approval-price correlations may be explained by the known approval rating decrease over a presidency's course \cite{beck2012what}, coupled with the trend of increased asset prices over time \cite{jorda2019rate}.
Fitting for a memecoin, \$DOGE is an exception: its correlation is inconsistent, sometimes weakly negative or positive (as with all administrations besides Trump's first), and at other times moderately negative (under Trump's first administration).

\paragraphNoSkip{Trump Effect II: General Token Market}
The correlation of approval with log-returns (that is, the log of the ratio between two consecutive price points) reveals that Trump's second term represents a marked departure from prior administrations.
Across assets, consistent negative associations are observed: \$BTC ($\rho=-0.2$), \$ETH ($\rho=-0.33$), \$SOL ($\rho=-0.24$), \trump and \melania (both $\rho=-0.19$).
This does not hold for past administrations.

\paragraphNoSkip{Comparing Correlations}
For \trump and \melania, the negative correlation between approval ratings and log-returns contrast with the positive one for token prices.
That is because a positive correlation with prices does not imply a correlation with price \emph{increases}, but rather with a high price \emph{level}.
That is, while returns may be negative, as long as the price level remains relatively high, a positive correlation with prices is possible.
Next, we wish to dwell on the economic implications of this finding.
To reiterate, the data indicate that high presidential approval ratings correspond to \emph{high prices} and thus also to \emph{negative returns}.
This is suggestive of mean-reversion: high price levels correspond to a decrease in prices.
Observed differently, these correlations possibly align with the literature on the mean-reversion of approval ratings \cite{lebo2007aggregated, jr1992error}.

\paragraphNoSkip{\gls{ADF} Test}
The test reveals that \trump prices are non-stationary, while returns are stationary. This is standard for asset prices.

\subsection{Event Study}
We apply the \gls{PELT} algorithm to the log-returns and prices of \trump and \melania to identify significant structural breaks from in the time series.
By matching these breaks with the corpus' events for exactly the same day, we identify several key change points that align with major $2025$ events (see \cref{fig:Timeline}):

\paragraphNoSkip{Feb. 1}
Announcement of $25\%$ Mexico and Canada tariffs.

\paragraphNoSkip{Apr. 2}
Broad $10\%$ tariffs announced (``Liberation Day'').

\paragraphNoSkip{Apr. 23}
``Crypto Dinner'' announced for top \trump holders.

\paragraphNoSkip{May 30}
Musk farewell ceremony at the Oval Office.

\paragraphNoSkip{Jun. 17}
Passage of crypto-friendly \gls{GENIUS} Act.

\paragraphNoSkip{Sep. 21}
Charlie Kirk's memorial.

\paragraphNoSkip{Oct. 10}
Trump threats China with $130\%$ tariffs.

\subsection{Ecosystemic Impact}
The Jan. 17 launch of \trump had a measurable impact on Solana, the blockchain on which it launched, and on the ecosystem at large.
We now elaborate on both.

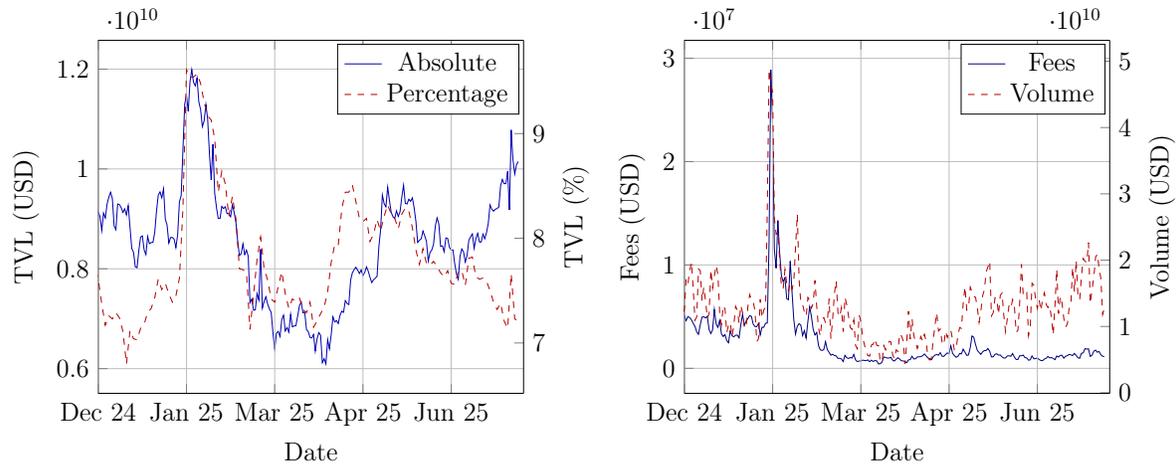
\begin{figure*}
  \centering
  \tikzsetnextfilename{SolanaTvlShareTvl}
  \begin{minipage}{.48\textwidth}
  \centering
  \scalebox{0.825}{\begin{tikzpicture}
    \begin{axis}[
      xlabel={Date},
      ylabel={TVL (USD)},
      xmajorgrids,
      ymajorgrids,
      date coordinates in=x,
      xmin=2024-12-01,
      xmax=2025-07-30,
      axis x line=none,
      axis y line*=left,
      axis x line*=bottom,
    ]
      \addplot [draw=blue!70!black] table [col sep=comma, x=snapped_at, y=TVL] {.private/data/DefiLlama/sol_data.csv};
      \label{fig:SolanaTvlShareTvl}
    \end{axis}

    \begin{axis}[
      xlabel={Date},
      ylabel={TVL (\%)},
      xmajorgrids,
      date coordinates in=x,
      xmin=2024-12-01,
      xmax=2025-07-30,
      xticklabel={\pgfcalendar{tickcal}{\tick}{\tick}{\pgfcalendarshorthand{m}{.}} \short{\year}},
      axis y line*=right,
      legend style={at={(0.99,0.99)}},
    ]
      \addlegendimage{/pgfplots/refstyle=fig:SolanaTvlShareTvl};
      \addlegendentry{Absolute};

      \addplot [dashed, draw=red!70!black] table [col sep=comma, x=snapped_at, y=Solana] {.private/data/DefiLlama/TVL/chains_tvl.csv};
      \addlegendentry{Percentage};
    \end{axis}
  \end{tikzpicture}}
  \end{minipage}
  \hfill
  \tikzsetnextfilename{SolRevenueFeesTvlFees}
  \begin{minipage}{.48\textwidth}
  \centering
  \scalebox{0.825}{\begin{tikzpicture}
    \begin{axis}[
      xlabel={Date},
      ylabel={Fees (USD)},
      xmajorgrids,
      ymajorgrids,
      date coordinates in=x,
      xmin=2024-12-01,
      xmax=2025-07-30,
      axis x line=none,
      axis y line*=left,
      axis x line*=bottom,
    ]
      \addplot [draw=blue!50!black] table [col sep=comma, x=snapped_at, y=Chain Fees] {.private/data/DefiLlama/sol_data.csv};
      \label{fig:SolRevenueFeesTvlFees}
    \end{axis}
    \begin{axis}[
      xlabel={Date},
      ylabel={Volume (USD)},
      xmajorgrids,
      date coordinates in=x,
      xmin=2024-12-01,
      xmax=2025-07-30,
      xticklabel={\pgfcalendar{tickcal}{\tick}{\tick}{\pgfcalendarshorthand{m}{.}} \short{\year}},
      axis y line*=right,
      legend style={at={(0.99,0.99)}},
    ]
      \addlegendimage{/pgfplots/refstyle=fig:SolRevenueFeesTvlFees};
      \addlegendentry{Fees};
      \addplot [dashed, draw=red!70!black] table [col sep=comma, x=snapped_at, y=DEXs Volume] {.private/data/DefiLlama/DEX/dexs_vol.csv};
      \addlegendentry{Volume};
    \end{axis}
  \end{tikzpicture}}
  \end{minipage}
  \caption{Solana's share of all \gls{DeFi} \gls{TVL} spikes to a record 9.61\% (left figure), and both its daily fees and \glspl{DEX} trading volume surge to their all-time highs shortly after the \trump launch on Jan. 17 (right figure).}
  \label{fig:SolanaImpact}
\end{figure*}

\paragraphNoSkip{Impact on Solana}
As \cref{fig:SolanaImpact} (right) shows, the launch was followed by immediate and massive trading activity on Solana-based \glspl{DEX}, with volume reaching an all-time high of nearly $\$50$ billion.
This intense activity is naturally accompanied by spikes in other metrics as well, with daily transaction fees peaking at $\$28$ million.

\paragraphNoSkip{Impact on Cryptoasset Market}
The launch coincided with liquidity flow to the Solana ecosystem: \cref{fig:SolanaImpact} (left) shows that Solana's \gls{DeFi} \gls{TVL} (comprising both the native token and all on-chain assets) experienced a sharp increase, rising from approximately $\$8.5$ billion to a peak of $\$12$ billion in late January.
Similarly, Solana's share of the broad crypto-market \gls{TVL} (i.e., the total \gls{DeFi} \gls{TVL} across all blockchains) spiked to a record of $9.61\%$ in the week following the launch.
This increase implies that capital migrated from competing blockchains (e.g., Ethereum) towards Solana.

\section{Conclusion}
\label{sec:Conclusion}
We empirically analyze the emerging \gls{PoliFi} ecosystem.
Our focus is on two prominent representatives, the \trump and \melania tokens, covering both from their launch to the time of writing, resulting in a preliminary analysis of their first 10 months.
We identify and quantify the ``Trump Effect'', which comes to the fore in two ways: in the positive correlation between ratings and the prices of both \trump and \melania, and in the generally consistent negative correlation between approval ratings and token returns.
In addition, we find preliminary evidence of inefficiencies in the \gls{PoliFi} market, with canonical methods indicating that \trump prices exhibit long-term memory.
Finally, we quantify the ecosystemic impact of the launch of \trump, demonstrating that the launch coincided with a short-term boost to Solana across several metrics, e.g., \gls{TVL} market share.
In total, we provide an empirical perspective on how public perception of political figures affects the corresponding \gls{PoliFi} assets.

The current study is a work-in-progress: while we analyze data covering the entire existence of \trump and \melania, additional data is required to reach stronger conclusions.
Thus, our work can be seen as shedding light on some aspects of the nascent and complex \gls{PoliFi} ecosystem.
As these assets mature, we will continue to update our study with both additional data and analyses.

\printbibliography[heading=bibintoc]

\begin{figure}
  \centering
  \tikzsetnextfilename{TrumpT2ApprovalVsTokens}
  \scalebox{0.99}{\begin{tikzpicture}
    \begin{axis}[
      ylabel={Approval (\%)},
      ymajorgrids,
      axis x line=none,
      date coordinates in=x,
      xmin=2025-01-20,
      xmax=2025-10-29,
      axis y line*=left,
      axis x line*=bottom,
    ]   
      \addplot [draw=blue, style=densely dashdotdotted] table [col sep=comma, x=snapped_at, y=pct] {figures/TrumpT2_all_tokens.csv};
      \label{fig:ApprovalApprLegend}
    \end{axis}
    \begin{axis}[
      xlabel={Date},
      ylabel={Log Log Price (USD)},
      xmajorgrids,
      axis y line*=right,
      date coordinates in=x,
      xmin=2025-01-20,
      xmax=2025-10-29,
      xticklabel style={rotate=90, anchor=near xticklabel,},
      legend style={at={(0.99, 0.6)}, nodes={scale=0.65, transform shape}, fill=white, fill opacity=0.6, text opacity=1},
      xticklabel={\pgfcalendar{tickcal}{\tick}{\tick}{\pgfcalendarshorthand{m}{.}} \short{\year}},
    ]
      \addlegendimage{/pgfplots/refstyle=fig:ApprovalApprLegend};
      \addlegendentry{Approval}; %
      
      \addplot [draw=orange!70!black] table [col sep=comma, x=snapped_at, y=TRUMP_price, y expr=ln(\thisrow{log_TRUMP_price})] {figures/TrumpT2_all_tokens.csv};
      \addlegendentry{\trump};

      \addplot [draw=green!50!black] table [col sep=comma, x=snapped_at, y=BNB_price, y expr=ln(\thisrow{log_BNB_price})] {figures/TrumpT2_all_tokens.csv};
      \addlegendentry{\$BNB};

      \addplot [draw=red!70!black] table [col sep=comma, x=snapped_at, y=SOL_price, y expr=ln(\thisrow{log_SOL_price})] {figures/TrumpT2_all_tokens.csv};
      \addlegendentry{\$SOL};

      \addplot [draw=yellow!70!black] table [col sep=comma, x=snapped_at, y=DOGE_price, y expr=ln(\thisrow{log_DOGE_price})] {figures/TrumpT2_all_tokens.csv};
      \addlegendentry{\$DOGE};

      \addplot [draw=purple!70!black] table [col sep=comma, x=snapped_at, y=ETH_price, y expr=ln(\thisrow{log_ETH_price})] {figures/TrumpT2_all_tokens.csv};
      \addlegendentry{\$ETH};
    \end{axis}
  \end{tikzpicture}}
  \caption{\trump's price follows approval, with the notable exception of a 3-week period starting with ``liberation day''. Other notable tokens do not exhibit this behavior.}
  \label{fig:TrumpT2ApprovalVsTokens}
\end{figure}
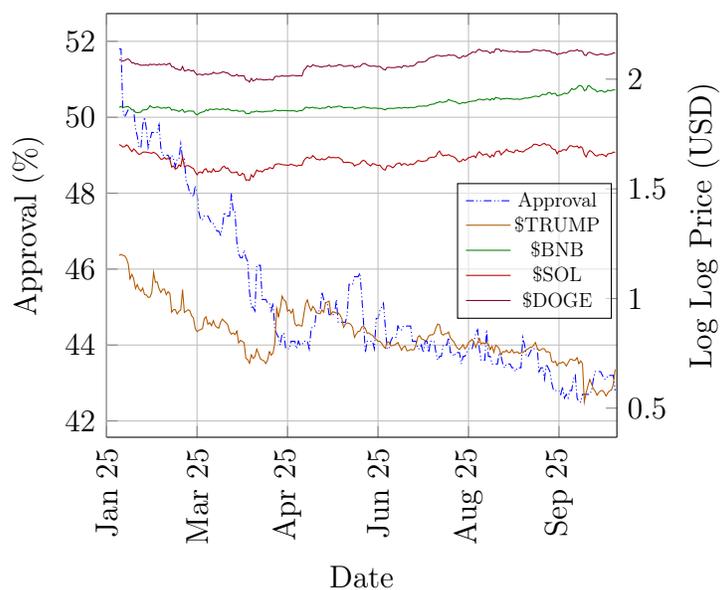

\appendix

\newpage
\section{Glossary}
\label[appendix]{sec:Glossary}
Following is a list of the notations and acronyms used in the paper.
\setglossarystyle{alttree}\glssetwidest{AAAAAAAA} %
\printnoidxglossary[type={symbols}]
\printnoidxglossary[type={acronym}]
\end{document}